\documentclass[10pt, a4paper]{article} 

\usepackage{indentfirst}

\usepackage{xcolor}

\usepackage{ulem}
\usepackage{amsthm}
\usepackage{marginnote}

\usepackage[english]{babel}
\usepackage{graphicx} 

\usepackage{cite}

\usepackage[thinc]{esdiff}

\usepackage{tabularx}

\usepackage{latexsym}
\usepackage{amsfonts} 
\usepackage{amsmath} 

\usepackage{multirow}

\usepackage{enumitem}

\usepackage[utf8]{inputenc} 
\usepackage[T1,T2A]{fontenc}

\usepackage{layout}

\usepackage[left]{lineno}

\usepackage{color, soul} 
\setulcolor{yellow}

\usepackage[colorlinks=true, citecolor=blue]{hyperref}

\usepackage[nameinlink]{cleveref} 

\newtheorem{theorem}{Theorem}
\newtheorem*{theorem*}{Theorem}

\newtheorem{cor}{Corollary}

\newtheorem*{note*}{Remark}
\newtheorem*{def*}{Definition}

\title{Implementing Dynamic Pricing Across Multiple Pricing Groups in Real Estate}

\author{Lev Razumovskiy \and Mariya Gerasimova \and Nikolay Karenin \and Mikhail Safro}
\newcommand{\Addresses}{{
\bigskip
\footnotesize

Lev Razumovskiy	, \textsc{RAMAX Group}\par\nopagebreak
  \textit{E-mail address} : \text{lev.razumovskiy@ramax.com}
  
\medskip
  
 Mariya Gerasimova \textsc{}\par\nopagebreak
  \textit{E-mail address} : \text{marijyagerasimova@yandex.com}
  
\medskip
Nikolay Karenin, \textsc{RAMAX Group}\par\nopagebreak
  \textit{E-mail address} : \text{nikolay.karenin@ramax.com}

\medskip
Mikhail Safro, \textsc{RAMAX Group}\par\nopagebreak
  \textit{E-mail address} : \text{msafro@ramax.com}

}}
\date{}

\begin{document}
\maketitle

\begin{abstract}
This article presents a mathematical model of dynamic pricing for real estate (RE) that incorporates multiple pricing groups, thereby expanding the capabilities of existing models. 
The developed model solves the problem of maximizing aggregate cumulative revenue at the end of the sales period while meeting the revenue and sales goals.
A method is proposed for distributing aggregate cumulative revenue goals across different RE pricing groups. 
The model is further modified to account for the time value of money and the real estate value increase as construction progresses.
The algorithm for constructing a pricing policy 
for multiple pricing groups 
is described, and numerical simulations are performed to demonstrate how the algorithm operates.
\end{abstract}

\section*{Introduction}
Dynamic pricing is one of the key topics in economics and management, as it provides flexible methods for adjusting prices based on various factors such as demand, competition, and changes in central bank interest rates.
This approach enables companies not only to manage resources effectively and optimize profits, but also to respond to market changes promptly, thus enhancing their competitiveness.

Currently, dynamic pricing models are developed and applied across various sectors of the economy.
For example, in air transportation~\cite{Fiig-18}, dynamic pricing is used to manage fares; in the hotel industry~\cite{Cho-18, Bayoumi-13}, it helps to optimise room rates; and in taxi services~\cite{Qian-17}, it allows for real-time price updates throughout the day.

For development companies, dynamic pricing is also a crucial business management tool. 
Real estate has unique characteristics that distinguish this field from others. 
For instance, the product being sold --- RE objects -- is finite, irreplaceable, and must be fully sold by the end of the sales period. 
It is worth noting that the sales period for real estate is significantly longer compared to, for example, retail~\cite{Riseth-19, Benmark-17}, and can extend over several years.
Among the most significant studies on dynamic pricing in real estate are the foundational works of Kinkaid and Darling~\cite{Kinkaid-62} and Gallego and van Ryzin~\cite{Gallego-13}, who were among the first to address the dynamic pricing problem with a fixed sales period and constraints on sales and revenue.
Further variations and modifications of the model can be found in other works~\cite{Feng-95, Bitran-97, Feng-00b}.
Additionally, the work of Besbes and Maglaras~\cite{Besbes-12} formulates the optimization problem of determining the optimal pricing policy and provides a solution for the case where total demand and conversion are known.
The work~\cite{Razumovsky-24} generalizes this model for the case of variable total demand, accounting for the time value of money and the RE objects value increase as construction progresses.

In the aforementioned studies, all units in a residential complex were assumed to be homogeneous; that is, the models did not differentiate between, for example, one-bedroom and two-bedroom apartments.
However, from a practical perspective, such an approach is unlikely to be satisfactory.
In this work, over a finite time interval \([0,T]\), we consider the problem of revenue optimization for the sale of a real estate complex, which is not homogeneous, but instead is divided into 
\(k\) pricing groups (\(k>1\)).
All units within a given pricing group have identical characteristics and prices, which may change over time individually.
We provide a detailed analysis of various types of constraints on sales and revenue, which naturally arise from business requirements.
Additionally, the paper examines the case of variable time value of money, which can be represented as a monotonically non-increasing function of time (a characteristic induced by factors such as inflation, investment activity, etc.), and the RE objects value, which increases as construction progresses.

In Section~\ref{sec:base_model}, we describe the basic mathematical model for revenue maximization in the presence of \(k\) pricing groups, assuming that the demand for each pricing group depends only on the price of RE objects in that group, meaning there is no "cannibalization" process between pricing groups. 
We formulate and prove a result regarding the general form of the pricing policy, identifying the problem of revenue distribution across the pricing groups. 
An algorithm for determining a pricing policy for each pricing group is presented.
Section~\ref{sec:distribution} introduces a solution to the distribution problem.
Section~\ref{sec:tvm} generalizes the basic model to account for the time value of money and the RE objects value increase as construction progresses, and presents an algorithm for determining the pricing policy. 
In Section~\ref{sec:sims}, the basic algorithm is demonstrated for the case of two real estate pricing groups: one- and two-bedroom apartments. 
We then propose an alternative approach to distributing aggregate cumulative revenue plans across pricing groups, followed by a comparison with the previously presented method. 
The algorithm’s operation is demonstrated in a scenario where demand changes for one of the pricing groups.

The main contribution of this work lies in the development of a mathematical model that incorporates multiple real estate pricing groups, each with its own pricing policy and demand function.
This model allows us to examine how prices change with time and in response to demand changes, considering revenue and sales constraints for each pricing group.
Furthermore, this article proposes a modification of the basic model to include the time value of money and the RE objects value increase as construction progresses, making the model more 
applicable to
real-world market conditions.

\section{Mathematical model of pricing for RE objects of different pricing groups} \label{sec:base_model}

In this section we describe the basic dynamic pricing model for the case of several pricing groups, and formulate the basic task of finding the optimal pricing policy.


We introduce the following notation: 
\begin{itemize}
    \item[] $p^i(t)$ -- the pricing policy for an RE object of pricing group $i$; 
    \item[] $v^i(p)$ -- the demand function of pricing group $i$. This quantity represents how many RE objects of pricing group $i$ would be purchased per unit of time if the set price is $p$.
\end{itemize}

The revenue rate from the sale of RE of pricing group \(i\) with price \(p^i\) and sales rate \(v^i(p^i)\) is equal to \(p^i \cdot v^i(p^i)\). Then, the cumulative sales \(S^i_{p^i(t)}(t)\) and cumulative revenue \(R^i_{p^i(t)}(t)\) from selling all RE objects in pricing group \(i\) over the time period from \(0\) to \(t\) are expressed as follows:
\begin{eqnarray*}
    S^i_{p^i(t)}(t) = S^i(t) = \int_0^t v^i(p^i(\tau)) \, d\tau,\\
    R^i_{p^i(t)}(t) = R^i(t) = \int_0^t p^i(\tau) \, v^i(p^i(\tau)) \, d\tau.
\end{eqnarray*}

The cumulative revenue from sales of all RE objects of pricing group \(i\) over the entire sales period \(T\) is given by:
\begin{equation*}
    R^i_{p^i(t)}(T) = R^i(T) = \int_0^T p^i(t) \, v^i(p^i(t)) \, dt.
\end{equation*}

However, in practical applications~\cite{Faruqui-10, Saharan-20, Mcafee-06}, the focus is not on the revenue from selling RE objects within a specific pricing group, but rather on the total cumulative revenue from selling RE objects across all pricing groups. Therefore, the quantity of interest is
\begin{equation*}
    R_{p^\bullet(t)}(t) = R(t) = \sum_{i=1}^k R^i_{p^i(t)}(t),
\end{equation*}
which is referred to as the \textit{aggregate cumulative revenue} in this work.

We will also introduce notation for cumulative sales and cumulative revenue for each pricing group as well as for the aggregate cumulative revenue accumulated over the interval \([t_1, t_2] \subset [0, T]\):
\begin{eqnarray*}
    S^i_{p^i(t)}(t_1, t_2) = S^i(t_1, t_2) = \int_{t_1}^{t_2} v^i(p^i(t)) \, dt,\\
    R^i_{p^i(t)}(t_1, t_2) = R^i(t_1, t_2) = \int_{t_1}^{t_2} p^i(t) \, v^i(p^i(t)) \, dt,\\
    R_{p^\bullet(t)}(t_1, t_2) = R(t_1, t_2) = \sum_{i=1}^k R^i(t_1, t_2).
\end{eqnarray*}

Let us denote by
\begin{eqnarray*}
    R^{*i}_{p^i_0}(t_1, t_2) = R^{*i}(t_1, t_2) = p^i_0 \cdot v^i(p^i_0) \cdot (t_2 - t_1),\\
    R^*_{p^\bullet_0}(t_1, t_2) = R^*(t_1, t_2) = \sum_{i=1}^k R^{*i}(t_1, t_2)
\end{eqnarray*}
the cumulative revenue for pricing group \(i\) and the aggregate cumulative revenue accumulated over the interval \([t_1, t_2] \subset [0, T]\) under constant prices \(p^1(t) = p^1_0, \ldots, p^k(t) = p^k_0\). For convenience, we also introduce the notation: 
$$R^{*i}_{p^i_0}(t) = R^{*i}(t) = R^{*i}_{p^i_0}(0, t),$$ 
$$R^{*}_{p^\bullet_0}(t) = R^{*}(t) = \sum_{i=1}^k R^{*i}_{p^i_0}(t).$$

In the following, we will implicitly assume that the pricing policy \((p^1(t), \ldots, p^k(t))\) is clear from the context and omit the subscript in the formulas.

Additionally, when determining pricing policy, it is necessary to take into account several constraints on both the total quantity of RE objects sold within each pricing group and the aggregate cumulative revenue from the sales of RE objects across all available pricing groups.

Without loss of generality, we will assume that the constraints on sales and aggregate cumulative revenue are set at the same fixed moments in time \(0 = \tau_0 < \tau_1 < \ldots < \tau_l = T\). Let \(S^i_j\) denote the sales constraint for the \(i\)-th pricing group at time \(\tau_j\), and \(R_j\) denote the aggregate cumulative revenue constraint at time \(\tau_j\). Therefore, the problem of finding the optimal pricing policy in the context of multiple pricing groups can be formulated as follows:
\begin{equation}\label{eq:total_rev}
    R(T) \rightarrow \mathrm{max}
\end{equation}

subject to the following constraints:
\begin{itemize}
\item[---] on intermediate sales
\begin{equation} \label{constr:sales}
    S^i(\tau_j) \geq S^i_j,  \quad 1 \leq j \leq l-1, \; 1 \leq i \leq k,
\end{equation}
\item[---] on final sales
\begin{equation} \label{constr:end_sales}
    S^i(T) = S^i_l, \quad 1 \leq i \leq k,
\end{equation}
\item[---] on aggregate cumulative revenue
\begin{equation} \label{constr:revenue}
    R(\tau_j) \geq R_j, \quad 1 \leq j \leq l.
\end{equation}
\end{itemize}

\subsection{General form of the optimal 
pricing policy}

In this section, we explore the solution to the problem~(\ref{eq:total_rev})--(\ref{constr:revenue}). First, we consider a simpler scenario with constraints only on the final sales for each pricing group. We also formulate and prove the corresponding theorem.

\begin{theorem} \label{theorem:uniform_sales}
    The solution to problem~(\ref{eq:total_rev}), (\ref{constr:end_sales}) for each \(i = 1, \ldots, k\) is the constant prices \(p^i(t) = p^i_0\) for all \(t \in [0, T]\), such that
    \begin{eqnarray*}
        v^i(p^i_0) = \frac{S^i_l}{T}.
    \end{eqnarray*}
\end{theorem}

\begin{proof}
    Since for each \(i = 1, \ldots, k\) the revenue \(R^i(T)\) for each pricing group is a positive quantity, the expression
    $$
    R(T) = \sum_{i=1}^k R^i(T) = \sum_{i=1}^k \int_0^T p^i(t) v^i(p^i(t)) dt
    $$
    reaches its maximum if and only if each \(R^i(T)\) is maximized.

    Now, because the constraints~(\ref{constr:end_sales}) apply independently, we have a separate optimization problem for each pricing group. Thus, for each \(i\) from \(1\) to \(k\), the pricing policy is found as the solution to the following problem:
    \begin{eqnarray*}
        \int_0^T p^i(t) v^i(p^i(t)) dt \to \mathrm{max},
    \end{eqnarray*}
    subject to
    \begin{eqnarray*}
        \int_0^T v^i(p^i(t)) dt = S^i_l.
    \end{eqnarray*}

    According to~\cite{Besbes-12}, the solutions to these problems are constant prices \((p^1_0, \ldots, p^k_0)\) over the entire interval, which satisfy the equations:
    \begin{equation*}
        v^i(p^i_0) = \frac{S^i_l}{T}, \quad i = 1, \ldots, k.
    \end{equation*}
\end{proof}

The proven statement essentially means that, in the absence of additional constraints, the optimal strategy is an even (inventory) absorption.


\begin{figure}[!h]
    \includegraphics[width=1\linewidth]{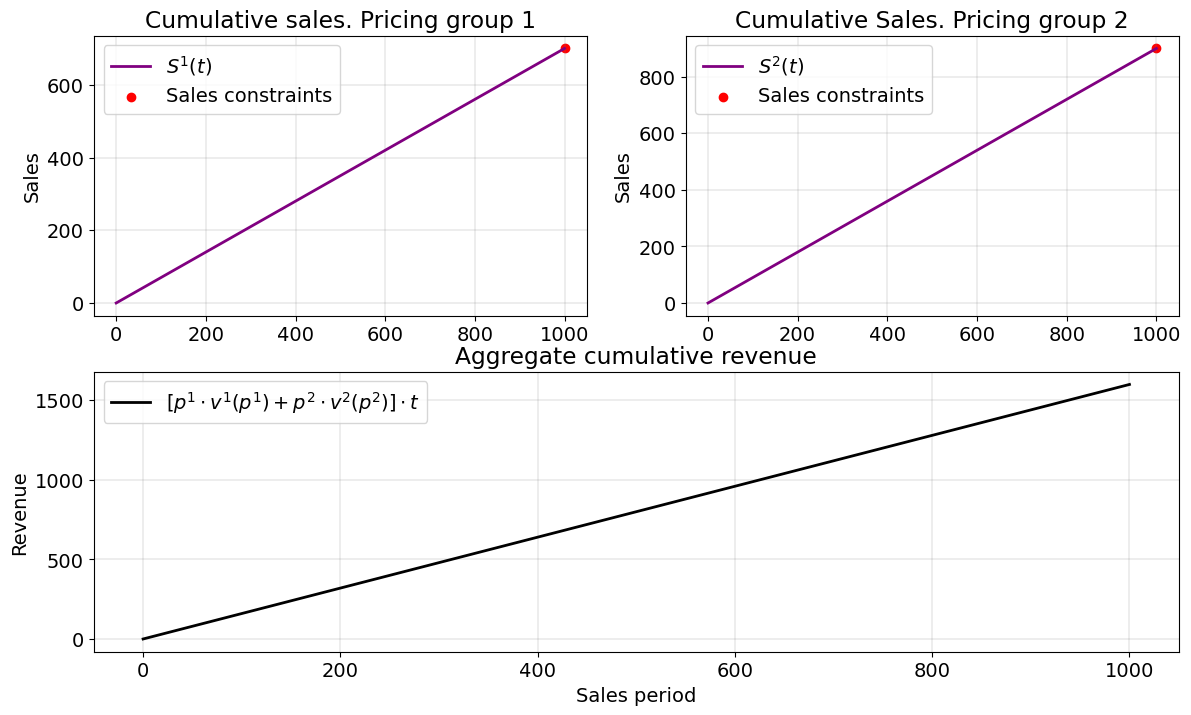}
    \caption{Even absorption for each pricing group and the corresponding aggregate cumulative revenue}
    \label{fig:uniform_sales}
\end{figure}

In Figure~\ref{fig:uniform_sales}, the statement of Theorem~\ref{theorem:uniform_sales} is illustrated for the case of two pricing groups (\(k=2\)). The top two graphs show the cumulative sales curve for each pricing group. The graph is a straight line, reflecting that the RE objects are being absorbed evenly across each pricing group. The lower revenue graph also shows that with even absorption, the aggregate cumulative revenue is accumulated uniformly.

The solution to the general problem, when additional constraints on intermediate sales and aggregate cumulative revenue are imposed, can now be derived as a corollary of Theorem~\ref{theorem:uniform_sales}.

\begin{cor} \label{cor:general}
    Let \((p^1(t), \ldots, p^k(t))\) be a solution to problem~(\ref{eq:total_rev})--(\ref{constr:revenue}). Then, for \(i = 1, \ldots, k\), the pricing policy \(p^i(t)\) is a piecewise constant function on the interval \([0, T]\), with value changes only occurring at the constraint times where they are strictly met.
\end{cor}

\begin{proof}
    Consider an arbitrary time interval \([\tau_{j-1}, \tau_j]\) between constraints. The corresponding constraint on sales for pricing group \(i\) will be given by the inequality \(S^i(\tau_j) \geq S^i_j\), while the constraint on aggregate cumulative revenue is given by the inequality \(\sum_{i=1}^k R^i(\tau_j) \geq R_j\). Initial conditions for the problem~(\ref{eq:total_rev})-(\ref{constr:revenue}) are formulated as: \(\forall i = 1, \ldots, k \; S^i(0) = 0, R^i(0) = 0\). By changing the variable \(t \mapsto t - \tau_{j-1}\) and introducing new functions \(\widetilde{S}^i, \widetilde{R}^i\) such that \(\widetilde{S}^i(t) = S^i(\tau_{j-1}) + S^i(t+\tau_{j-1})\) and \(\widetilde{R}^i(t) = R^i(\tau_{j-1}) + R^i(t+\tau_{j-1})\), we obtain, on the interval \([\tau_{j-1}, \tau_j]\), the same optimization problem as (\ref{eq:total_rev}), (\ref{constr:end_sales}), given the fixed final sales \(\widetilde{S}^i(\tau_{j}-\tau_{j-1})\). Therefore, by Theorem~\ref{theorem:uniform_sales}, with constraints on intermediate sales, the price must be constant over the interval, so for \(t \in [\tau_{j-1}, \tau_j]\), the pricing policy \((p^1(t), \ldots, p^k(t))\) will contain of constant prices.

    It remains to show that changes in pricing policy only occur at the strictly met constraints. For any \(i = 1, \ldots, k\), consider the pricing policy \(p^i(t)\). Suppose the pricing policy \(p^i(t)\) changes at time \(\tau_j\). On the interval \([\tau_{j-1}, \tau_{j+1}]\) between constraints, we fix the prices of the other pricing groups and the sales for the pricing group $i$. We obtain the problem for a single pricing group as discussed in~\cite{Besbes-12}, where it is already shown that price changes occur only at constraints that are strictly met, and that the optimal pricing policy does not decrease when it changes.
\end{proof}

\subsection{Features of constraints on aggregate cumulative revenue in the problem for multiple pricing groups} \label{sec:subsec:distrib_problem_statement}

Corollary~\ref{cor:general} only indicates the form that the optimal pricing policy should take and the times at which it may change. However, it does not specify what the pricing policy should actually be. In~\cite{Besbes-12}, the problem of revenue optimization was studied assuming that all RE objects are homogeneous, i.e., when there is only one pricing group (\(k=1\)). In this situation, the constraint time was chosen as the time of price change if the constraint was the \textit{most stringent} (in which the inequality constraints are strict) across all constraints from that moment to the end of the sales interval.

The main property of the most stringent constraints in the single pricing group case was that, first, price changes occurred there, and second, the previous price change magnitude could be calculated based on it. In this section, we will also highlight those special constraints across all the constraints that allow us to determine the time and magnitude of price changes.

\begin{def*}
    Let \(p^1(t), \ldots, p^k(t)\) be the solution to problem~(\ref{eq:total_rev}), (\ref{constr:end_sales}). \textit{Burdensome} constraints are those constraints on aggregate cumulative revenue for which the following inequality holds:
    \begin{equation}\label{eq:burdensome}
        R^*(t) < R_j.
    \end{equation}
\end{def*}

When burdensome constraint exists at a specific time \(\tau_j\), revenue at \(\tau_j\) must be higher than it would be under even absorption. Note also that adding non-burdensome revenue constraints does not change the solution under constraints on final sales; it will still be given by even absorption and, moreover, by the same formula.

\begin{def*}
    Let \(j = 1, \ldots, l\) and \(\tau_{j_c}\) be the current time. Suppose there exists \(j_c < j \leq l\) such that \(R(\tau_{j_c}) + R^*(\tau_{j_c}, \tau_j) < R_j\); the set of such indices is denoted by \(J\). Then the constraint \(R_{j_*}\), where \(j_* = \arg \max_{j \in J} \frac{R_j - R(\tau_{j_c}) - R^*(\tau_{j_c}, \tau_j)}{\tau_j - \tau_{j_c}}\), is called the most stringent constraint, and \(\tau_{j_*}\) is the time of the most stringent constraint.
\end{def*}

This means that the most stringent constraint demands the highest rate of revenue.

In the work~\cite{Razumovsky-24}, an exact algorithm is provided for finding the optimal pricing policy for homogeneous RE objects. However, in the case of multiple pricing groups, a similar algorithm is unlikely to exist. To illustrate this, let us consider the case of two pricing groups and a single burdensome constraint with value \(R\) at time \(\tau \in (0, T)\). Let the corresponding demand functions be denoted by \(v^1(p^1)\) and \(v^2(p^2)\).

Since there is only one burdensome constraint, it is the most stringent one, and according to Corollary~\ref{cor:general}, the optimal pricing policies are piecewise constant functions \(p^1(t), p^2(t)\) with value changes at the constraint point. Thus, for \(i = 1, 2\),
\begin{equation*}
    p^i(t) =
    \begin{cases}
        p^i_1, & \text{if} \; t \in [0, \tau],\\
        p^i_2, & \text{if} \; t \in (\tau, T].
    \end{cases}
\end{equation*}

Then, the solution to the optimization problem~(\ref{eq:total_rev}), (\ref{constr:end_sales}), (\ref{constr:revenue}) is a pricing policy \((p^1(t), p^2(t))\) that maximizes the aggregate cumulative revenue
\begin{equation*}
    p^1_1 v^1(p^1_1) \tau + p^1_2 v^1(p^1_2) (T - \tau) + p^2_1 v^2(p^2_1) \tau + p^2_2 v^2(p^2_2) (T - \tau)
\end{equation*}
and satisfies the conditions
\begin{equation} \label{eq:conditions_sales}
    v^i(p^i_1) \tau + v^i(p^i_2) (T - \tau) = S^i_l, \quad i = 1, 2,
\end{equation}
\begin{equation} \label{eq:conditions_revenue}
    p^1_1 v^1(p^1_1) \tau + p^2_1 v^2(p^2_1) \tau = R.
\end{equation}

Now, using the method of Lagrange multipliers, we find the necessary optimality condition for the pricing policy \((p^1(t), p^2(t))\). We construct the Lagrangian
\begin{multline*}
    \mathcal{L}(p^1_1, p^1_2, p^2_1, p^2_2, q^1, q^2, q^3) = \biggl (p^1_1 v^1(p^1_1) \tau + p^1_2 v^1(p^1_2) (T - \tau) + p^2_1 v^2(p^2_1) \tau + p^2_2 v^2(p^2_2) (T - \tau) \biggr ) + \\
    + q^1 \biggl (v^1(p^1_1) \tau + v^1(p^1_2) (T - \tau) - S^1_l \biggr ) + q^2 \biggl (v^2(p^2_1) \tau + v^2(p^2_2) (T - \tau) - S^2_l \biggr ) + \\
    + q^3 \biggl (p^1_1 v^1(p^1_1) \tau + p^2_1 v^2(p^2_1) \tau - R \biggr ).
\end{multline*}

Thus, by setting the partial derivatives of \(\mathcal{L}\) equal to zero, we obtain for \(i=1,2\)
\begin{equation*}
    v^i(p^i_1) + p^i_1 \frac{dv^i}{dp^i}(p^i_1) + q^i \frac{dv^i}{dp^i}(p^i_1) + q^3 \biggl (v^i(p^i_1) + p^i_1 \frac{dv^i}{dp^i}(p^i_1) \biggr ) = 0,
\end{equation*}
\begin{equation*}    
    v^i(p^i_2) + p^i_2 \frac{dv^i}{dp^i}(p^i_2) + q^i \frac{dv^i}{dp^i}(p^i_2) = 0.
\end{equation*}

Therefore, even in this simplest case, determining the pricing policy requires solving a system of seven nonlinear equations. The complexity of the system and the number of equations increase with the number of pricing groups and the addition of intermediate constraints. Clearly, it is not possible to gain a solution to this system in a closed form. Therefore, in this work, we do not provide exact algorithms for finding the optimal pricing policy. Instead, we propose heuristic, or quasi-optimal, algorithms for determining the pricing policy.

Based on the previously introduced concepts and proven statements, we can outline an algorithm for finding the quasi-optimal pricing policy.

\begin{itemize}
    \item[1.] Set step \(m = 0\) and step index \(j_m = 0\). 
    \item[2.] At step \(m\), for each pricing group \(i\), find the price \(p^i_m\) that solves the corresponding problem for a single pricing group with only intermediate and final sales constraints.
    \item[3.] For each \(j > j_m\), find the aggregate cumulative revenue \(R^*(\tau_{j_m}, \tau_j) = R^*_{p^{\bullet}_m}(\tau_{j_m}, \tau_j)\) and form the set \(J_m = \{j \in (j_m, l] \; | \; R^*(\tau_{j_m}, \tau_j) < R_j\}\) of indices of burdensome constraints.
    \item[4.] If \(J_m\) is non-empty, i.e., if there are burdensome constraints on the interval \([\tau_{j_m}, T]\),
    \begin{itemize}
        \item[4.1.] then set \(j_*  = \arg \max_{j \in J_m} \frac{R(\tau_j) - R^*(\tau_{j_m}, \tau_j)}{\tau_j - \tau_{j_m}}\) -- the index of the most stringent constraint, \(R_{j_*}\) -- the most stringent constraint, and \(\tau_{j_*}\) -- the time of the most stringent constraint, i.e., the next price recalculation time; proceed to 5.,
        \item[4.2.] otherwise, set prices \(p^1_m, \ldots, p^k_m\) over the interval \([\tau_{j_m}, T]\) and end the algorithm.
    \end{itemize}
    \item[5.] Recalculate prices by solving the equation
    \begin{equation} \label{eq:stringent}
        \sum_{i=1}^{k} p^i_m \cdot v^i(p^i_m) = \frac{R_{j_*} - R^*(\tau_{j_m})}{\tau_{j_*} - \tau_{j_m}}
    \end{equation}
    \item[6.] Set the prices \((p^1_m, \ldots, p^k_m)\) found in 5. on the entire interval \([\tau_{j_m}, \tau_{j_*}]\).
    \item[7.] If \(\tau_{j_*} < T\), then \(m \leftarrow m+1\), \(j_m \leftarrow j_*\), proceed to 2.
\end{itemize}

\begin{figure}[!h]
    \centering
    \includegraphics[width=1\linewidth]{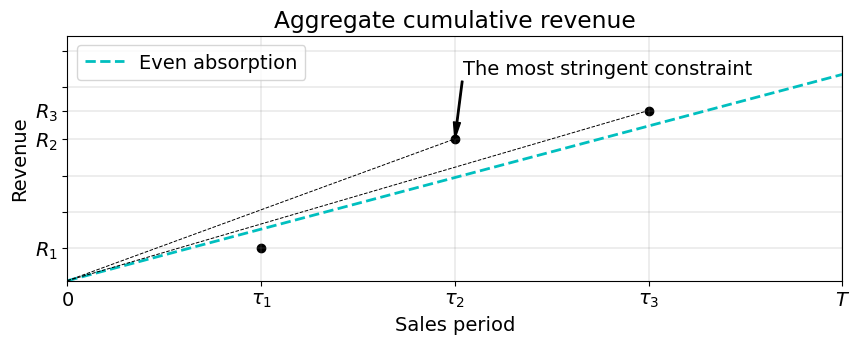}
    \caption{First step --- finding the most stringent constraint}
    \label{fig:burdensome_1}
\end{figure}

\begin{figure}[!h]
    \centering
    \includegraphics[width=1\linewidth]{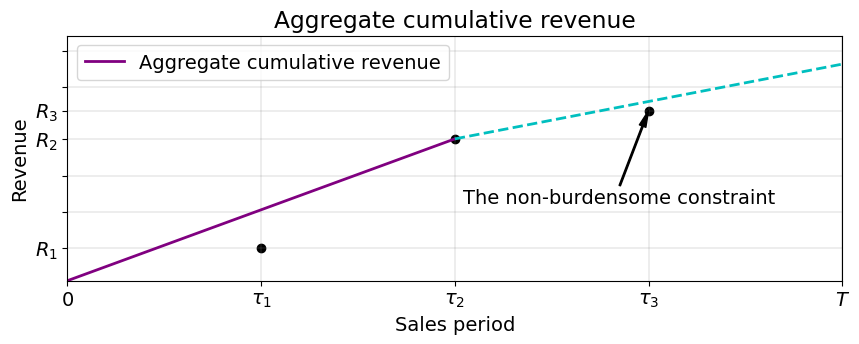}
    \caption{Second step --- approaching the constraint and identifying the next most stringent constraint}
    \label{fig:burdensome_2}
\end{figure}

\begin{figure}[!h]
    \centering
    \includegraphics[width=1\linewidth]{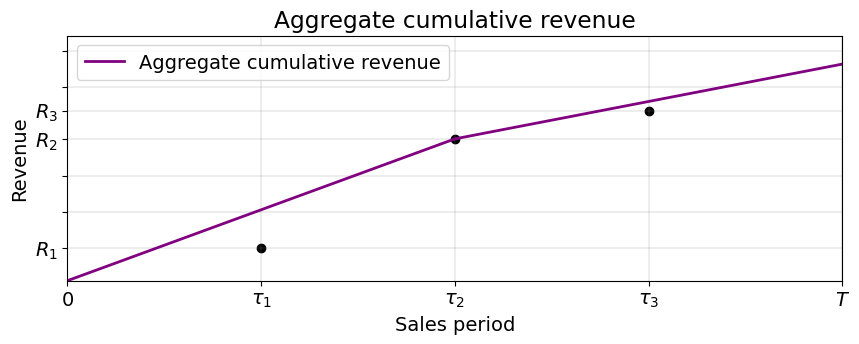}
    \caption{Third step --- aggregate cumulative revenue curve completed, algorithm finished}
    \label{fig:burdensome_3}
\end{figure}

We will now present a visual representation of the main part of this algorithm.
For simplicity, assume that sales constraints are only set at the end of the sales period at \(T=1000\), and constraints \(R_1, R_2, R_3\) on aggregate cumulative revenue are defined  at times \(\tau_1, \tau_2, \tau_3\). 
The algorithm’s operation is shown in Figures~\ref{fig:burdensome_1}-\ref{fig:burdensome_3}. 
In the first step (Fig.~\ref{fig:burdensome_1}), at time \(\tau = 0\)
prices are chosen for each pricing group according to item 2 of the algorithm. 
Then, the expected sales curve is calculated, shown as a dashed line connecting the points \((0, 0)\) and \((T, R^*(T))\). 
This curve corresponds to a situation in which RE objects are absorbed evenly (that is, most profitably from a business perspective) across each pricing group.

However, as it can be seen in the graph, two financial constraints, namely \(R_2\) and \(R_3\), cannot be met with even absorption. 
Therefore, moving to the second step (Fig.~\ref{fig:burdensome_2}), we identify the most stringent constraint among the burdensome ones \(R_2\) and \(R_3\). In the graph, the constraint \(R_2\) at time \(\tau_2\) is the most stringent, so, according to the algorithm, the minimal deviation from even absorption is made to satisfy this constraint.

Next, the new current time is set to \(\tau \leftarrow \tau_2\), and prices are determined again according to step 2. The expected sales curve (dashed line) is then recalculated. All remaining constraints on aggregate cumulative revenue can now be met with even absorption, so the algorithm leaves the prices as they are, producing the final aggregate cumulative revenue graph (Fig.~\ref{fig:burdensome_3}).

Unfortunately, as in the single pricing group case, obtaining a closed form expression for the unique solution of equation~(\ref{eq:stringent}) for multiple pricing groups is not possible. We call this problem the \textit{distribution problem}, and it is the focus of the next section.

\section{Distribution of constraints on aggregate cumulative revenue across pricing groups} \label{sec:distribution}

The exact solution to the distribution problem posed in the previous section is rather complex and requires solving a system of nonlinear equations. Since the demand functions \(v^1, \ldots, v^k\) are unknown in advance, and the concept of a demand function may only apply to relatively short time intervals, an approximate solution can be devised in the form of feedback-based algorithms relying on historical sales data. In this section, we will describe a potential approach to the distribution problem.

For convenience, let us introduce some notations. Let the step \(m\) of the algorithm and the pricing group number \(i \in \{1, \ldots, k\}\) be fixed, then
\begin{equation*}
    \Delta R_m := R(\tau_{j_*}) - R^*(\tau_{j_m}, \tau_{j_*}), \; \Delta \tau_m := \tau_{j_*} - \tau_{j_m}.
\end{equation*}
The value \(\Delta R_m\) indicates the extent to which the burdensome constraint \(R_{j_*}\) exceeds the revenue that could be achieved by even absorption from \(\tau_{j_m}\) to \(\tau_{j_*}\). Thus, \(\Delta R_m\) increases as the even absorption strategy becomes less applicable. The value \(\Delta \tau_m\) indicates only the duration between the current time \(\tau_{j_m}\) and the time of the burdensome constraint \(\tau_{j_*}\).

The distribution problem can be approached as follows. Revenue constraints will be allocated in proportion to the potential for generating more revenue, i.e., in proportion to the difference between the expected revenue achievable under even absorption and the maximum possible revenue. Thus, we finally reduce the problem of maximizing aggregate cumulative revenue for multiple pricing groups to a single group problem. We will now formalize this approach.

Let the prices \(p^1_m, \ldots, p^k_m\) be obtained at step \(m\) of the algorithm in step 2. Recall that the expected revenue gained over the interval \([\tau_{j_m}, \tau_{j_*}]\) for the \(i\)-th pricing group is
\begin{equation*}
R^{*i}(\tau_{j_m}, \tau_{j_*}) = p^i_m \cdot v^i(p^i_m) \cdot (\tau_{j_*} - \tau_{j_m}),
\end{equation*}
and the expected aggregate revenue at time \(\tau_{j_*}\) is given by
\begin{equation*}
    R^*(\tau_{j_m}, \tau_{j_*}) = \sum_{i=1}^k R^{*i}(\tau_{j_m}, \tau_{j_*}).
\end{equation*}
The maximum possible revenue for each pricing group \(i = 1, \ldots, k\) is determined by the formula
\begin{equation*}
    \widetilde{R}^i(\tau_{j_m}, \tau_{j_*}) :=   P^{i*}_m \cdot v^i(P^{i*}_m) \cdot (\tau_{j_*} - \tau_{j_m}),
\end{equation*}
where \(P^{i}_m = \arg \max p^i_m \cdot v^i(p^i_m)\). The aggregate maximum possible revenue, in turn, equals
\begin{equation*}
    \widetilde{R}(\tau_{j_m}, \tau_{j_*}) = \sum_{i=1}^k \widetilde{R}^i(\tau_{j_m}, \tau_{j_*}).
\end{equation*}

Let us denote by \(\eta^i_m\) the expression
\begin{equation*}
    \frac{\widetilde{R}^i(\tau_{j_m}, \tau_{j_*}) - R^{*i}(\tau_{j_m}, \tau_{j_*})}{\widetilde{R}(\tau_{j_m}, \tau_{j_*}) - R^*(\tau_{j_m}, \tau_{j_*})}.
\end{equation*}
It is evident that \(\forall i = 1, \ldots, k \; \eta^i_m \geq 0\) and \(\eta^1_m + \ldots + \eta^k_m = 1\).

Then, the portion of the revenue shortfall distributed to the \(i\)-th pricing group is defined by the following formula:
\begin{equation*}
    \Delta R^i_m = \eta^i_m \Delta R_m.
\end{equation*}

At step \(m\) of the algorithm in 5, we will recalculate the prices, as in the case of a single pricing group, using the equations
\begin{equation} \label{eq:distrib_prices}
    p^i_m \cdot v^i(p^i_m) = \frac{R^{*i}(\tau_{j_m}, \tau_{j_*}) + \Delta R^i_m}{\Delta \tau_m}.
\end{equation}

\section{The time value of money and the RE objects value increase as construction progresses} \label{sec:tvm}

Let us consider an enhancement to the basic model from Section~\ref{sec:base_model}, incorporating the time value of money and the RE objects value increase as construction progresses.
In the case of a single pricing group, the article~\cite{Razumovsky-24} presents a comparison of algorithm performance, assessing those that exclude these parameters and those that incorporate them.
It was shown that an algorithm that accounts for the time value of money yields higher revenue.
Building on these results for a single pricing group, in this section we will develop a modification of the current model, in which the time value of money and the RE value increase as construction progresses are incorporated.

Assume that the time value of money over the entire sales period \([0, T]\) is a non-increasing function of time \(\varphi(t)\).
The cumulative revenue for the \(i\)-th pricing group over the time period from \(0\) to \(t\) is given now by:
\begin{equation*}
    R^i(t) = \int_0^t \varphi(\tau)\cdot p^i(\tau) \cdot v^i(p^i(\tau)) d\tau.
\end{equation*}
In turn, the aggregate cumulative revenue considering the time value of money is as follows:
\begin{equation*}
    R(t) = \sum_{i=1}^k R^i(t) = \sum_{i=1}^k \int_0^t \varphi(\tau)\cdot p^i(\tau) \cdot v^i(p^i(\tau)) d\tau.
\end{equation*}

The following optimization problem can be posed:
\begin{equation} \label{eq:total_rev_tvm}
    R(T) = \sum_{i=1}^k \int_0^T \varphi(t)\cdot p^i(t) \cdot v^i(p^i(t)) dt \rightarrow \max,
\end{equation}
subject to constraints of the form~(\ref{constr:sales})--(\ref{constr:revenue}):
\begin{eqnarray*}
    S^i(\tau_j) \geq S^i_j, \quad 1 \leq j \leq l-1, \; 1 \leq i \leq k,\\
    S^i(T) = S^i_l, \quad 1 \leq i \leq k,\\
    R(\tau_j) \geq R_j, \quad 1 \leq j \leq l.
\end{eqnarray*}

When \(\varphi(t) \equiv 1\), i.e., when money does not change its value during the sales period, we arrive exactly at the base revenue optimization problem.

The value of RE objects depending on the construction stage can be considered a monotonically increasing function \(\kappa(t)\), which indicates how much more buyers are willing to pay for each property at time \(t\) compared to the beginning of sales. To incorporate this into our model, it is convenient to consider a family of demand functions \(v^i_t(p), i = 1, \ldots, k\), which depend on time \(t\), for which, for \(t \in [0, T) \), the following holds:
\begin{equation}
    v^i_t(\kappa(t)p^i) = v^i_0(p^i).
\end{equation}
It is convenient to use \(v^i(p) :=v^i_0(p), i = 1, \ldots, k\).
Let \(\widehat{p}^1(t), \ldots, \widehat{p}^k(t)\) be the pricing policy, and introduce auxiliary quantities \(p^i(t):=\frac{\widehat{p}^i(t)}{\kappa(t)}, i = 1, \ldots, k\). Then for any \(i = 1, \ldots, k\), we have
\begin{eqnarray}
    v^i_t(\widehat{p}^i(t)) = v^i_t(\kappa(t)p(t)) = v^i_0(p(t)) = v^i(p(t)),\\
    \varphi(t)\widehat{p}^i(t)v^i_t(\widehat{p}^i(t)) = \varphi(t)\kappa(t)p^i(t)v^i(p(t))=
    \zeta(t)p^i(t)v^i(p(t)),
\end{eqnarray}
where \(\zeta(t):=\varphi(t)\kappa(t)\) is the so-called generalized time value of money. 
We see that in this notation, the problem of maximizing the aggregate cumulative revenue at time $T$ considering \(\kappa(t)\) reduces to the previously examined problem, with \(\varphi(t)\) replaced by \(\zeta(t)\).
We will solve it in this exact form, using \(\varphi(t)\) instead of \(\zeta(t)\), for convenience.

Now, we describe the form that the solution to the above problem should take.
First, consider the case where there are only constraints on final sales.

\begin{theorem} \label{theorem:tvm}
    The solution to the problem~(\ref{eq:total_rev_tvm}), (\ref{constr:end_sales}) for each \(i = 1, \ldots, k\) is a pricing policy \(p^i(t)\) such that
    \begin{equation} \label{eq:price_eq_tvm}
        p^i(t) + \biggl[v^i(p^i(t))\biggr]\cdot\biggl[\frac{dv^i}{dp^i}(p^i(t))\biggr]^{-1} =-\frac{q^i}{\varphi(t)},
    \end{equation}
    where the numbers \(q^1, \ldots, q^k\) are obtained by substituting the pricing policy \(p^1(t), \ldots, p^k(t)\) into equations~(\ref{constr:end_sales}).
\end{theorem}

\begin{proof}
    As in the proof of Theorem~\ref{theorem:uniform_sales}, we have a problem with \(k\) pricing groups, which splits into \(k\) single pricing group problems \(i \in \{1, \ldots, k\}\):
    \begin{equation*}
        R^i(T) = \int_0^T \varphi(t) p^i(t) v^i(p^i(t)) dt \rightarrow \max,
    \end{equation*}
    subject to
    \begin{equation*}
        S^i(T) = \int_0^T v^i(p^i(t)) dt = S^i_l, \quad i = 1, \ldots, k.
    \end{equation*}
    Each problem for \(i\) is solved using the Lagrange method. The Lagrangian functional for each \(i = 1, \ldots, k\) is as follows:
    \begin{equation*}
        \mathcal{L}^i = \mathcal{L}^i(p^i(t); q^i) = \int_0^T \varphi(t) p^i(t) v^i(p^i(t)) dt + q^i \cdot \biggl (\int_0^T v^i(p^i(t)) dt - S^i\biggr ).
    \end{equation*}
    The necessary conditions for the pair \(p^i(t), q^i\) to be an extremum of the functional \(\mathcal{L}^i\) are two equations: \(\frac{\partial \mathcal{L}^i}{\partial p^i} = 0, \frac{\partial \mathcal{L}^i}{\partial q^i} = 0\). Thus, the solution to the problem~(\ref{eq:total_rev_tvm}), (\ref{constr:end_sales}) for each \(i = 1, \ldots, k\) is a function \(p^i(t)\) and a number \(q^i\) such that
    \begin{equation} \label{eq:tvm_solution_system}
        \begin{cases}
            \varphi(t) v^i(p^i(t)) + \varphi(t) p^i(t) \frac{dv^i}{dp}(p^i(t)) + q^i \cdot \frac{dv^i}{dp}(p^i(t)) = 0,\\
            \int_0^T v^i(p^i(t)) dt - S^i = 0.
        \end{cases}
    \end{equation}
    This finishes the proof.
\end{proof}

In general, it is not possible to obtain a closed-form solution to the system~(\ref{eq:tvm_solution_system}), as it represents a system of \(2k\) nonlinear equations. However, this is not necessary for practical application of the algorithm, since the solution only needs to be found for specific demand functions \(v^i\). For example, a linear function can be used to approximate the demand function:
\begin{equation} \label{eq:linear_el}
    v^i(p^i) = a^i - b^i \cdot p^i, \quad i = 1, \ldots, k,
\end{equation}
where the parameters \(a^i\) and \(b^i\) can be estimated from sales data.

In this particular case, the problem~(\ref{eq:total_rev_tvm}), (\ref{constr:end_sales}) can be solved explicitly. Substituting \(v^i(p^i(t)) = a^i - b^i \cdot p^i(t)\) into the first equation of the system~(\ref{eq:tvm_solution_system}) and finding \(q^i\) from the second equation, we obtain
\begin{equation} \label{eq:solution_tvm_linear}
        p^i(t) = \frac12 \biggl [ \frac{a^i}{b^i} - \frac{q^i}{\varphi(t)} \biggr ], 
\end{equation}
\begin{equation} \label{eq:constants}
    q^i = \frac{2S^i_l - a^i T}{b^i I(0, T)},
\end{equation}
where \(I(\tau_{(1)}, \tau_{(2)}) = \int_{\tau_{(1)}}^{\tau_{(2)}} \frac{dt}{\varphi(t)}\).

The obtained result holds in the absence of burdensome constraints~(\ref{eq:burdensome}). When such constraints exist, the constants \(q^i\) need to be adjusted.

The following result is a corollary of Theorem~\ref{theorem:tvm}.

\begin{cor}
    Let \((p^1(t), \ldots, p^k(t))\) be the solution to the problem~(\ref{eq:total_rev_tvm}), (\ref{constr:sales})--(\ref{constr:revenue}) with demand functions of the form~(\ref{eq:linear_el}). Then, for \(i = 1, \ldots, k\), the constants \(q^1, \ldots, q^k\) are piecewise constant functions on the interval \([0, T]\), and their values may only change at times when constraints are strictly met.
\end{cor}

\begin{proof}
    Consider an arbitrary time interval \([\tau_{j-1}, \tau_j]\) between constraints. 
    The corresponding constraint on sales for pricing group \(i\) is given by the inequality \(S^i(\tau_j) \geq S^i_j\), and the constraint on aggregate cumulative revenue is given by the inequality \(\sum_{i=1}^k R^i(\tau_j) \geq R_j\). The initial conditions in the problem~(\ref{eq:total_rev_tvm}), (\ref{constr:sales})--(\ref{constr:revenue}) are formulated as: \(\forall i = 1, \ldots, k \; S^i(0) = 0, R^i(0) = 0\). By changing the variable \(t \mapsto t - \tau_{j-1}\) and introducing new functions \(\widetilde{S}^i, \widetilde{R}^i\), such that \(\widetilde{S}^i(t) = S^i(\tau_{j-1}) + S^i(t+\tau_{j-1})\) and \(\widetilde{R}^i(t) = R^i(\tau_{j-1}) + R^i(t+\tau_{j-1})\), we have, on the interval \([\tau_{j-1}, \tau_j]\), the same optimization problem as (\ref{eq:total_rev_tvm}), (\ref{constr:end_sales}). According to Theorem~\ref{theorem:tvm} and formula~(\ref{eq:tvm_solution_system}), the constants \(q^1, \ldots, q^k\) will be constant on the interval \([\tau_{j-1}, \tau_j)\).

    Fix an arbitrary index \(i = 1, \ldots, k\) and consider the pricing policy \(p^i(t)\). Suppose the pricing policy \(p^i(t)\) changes at time \(\tau_j\). On the interval \([\tau_{j-1}, \tau_{j+1}]\) between constraints, we fix the prices of the other pricing groups and the sales for the given one. We obtain the problem for a single pricing group as discussed in~\cite{Besbes-12}, where it is already known that the constants \(q^1, \ldots, q^k\) only change at strictly met constraints.
\end{proof}

\textbf{Algorithm}
\begin{itemize}
    \item[1.] Set the current step \(m=0\), current step index \(j_m = 0\).
    \item[2.] Find the constants \(q^1_m, \ldots, q^k_m\) from the system~(\ref{eq:solution_tvm_linear}) on the interval \([\tau_{j_m}, T]\).
    \item[3.] For each \(j > j_m\), find the aggregate cumulative revenue \(R(\tau_{j_m}, \tau_j) = R_{p^{\bullet}_m}(\tau_{j_m}, \tau_j)\) where \(p^i_m\) is the price corresponding to the constant \(q^i_m\) (\(i = 1, \ldots, k\)). Then, form the set \(J_m = \{j \in (j_m, l] \; | \; R(\tau_{j_m}, \tau_j) < R_j\}\) of indices of constraints that were not met with the constants found at 2.
    \item[4.] Assume the elements in \(J_m\) are ordered, i.e., arrange them in ascending order and assign indices to them: \(j^{(1)} < \ldots < j^{(|J_m|)}\).
    \item[5.] For \(s = 1\), calculate the values of constants \(q^1_m, \ldots, q^k_m\) on the interval \([\tau_{j_m}, \tau_{j^{(1)}}]\) under the condition
        \begin{equation} \label{eq:constants_equation}
        \sum_{i=1}^k \int_0^{\tau_{j^{(s)}}} \frac{\varphi(t)}{4} \biggl [ \frac{a^i_m}{b^i_m} - \frac{q^i_m}{\varphi(t)} \biggr ] \biggl [ a^i_m + \frac{q^i_m b^i_m}{\varphi(t)} \biggr ] dt = R_{j^{(s)}}.
    \end{equation}
    \begin{itemize}
        \item[5.1.] Then, check if all constraints are satisfied with the current values of constants.
        \item[5.2.] If there is at least one unsatisfied constraint (with index \(j^{(s')}\)), change \(s\) to \(s'\) and recalculate the constants until all constraint are satisfied; else, perform the assignment \(j^* \leftarrow j^{(s)}\) and proceed to 6.
    \end{itemize}
    \item[6.] Set the pricing policy on the interval \([\tau_{j_m}, \tau_{j^*}]\) based on the system~(\ref{eq:tvm_solution_system}).
    \item[7.] If \(\tau_{j^*} < T\), then \(m \leftarrow m + 1\), \(j_m \leftarrow j^*\), proceed to 2.
\end{itemize}

Only step 5 remains unclear, which again leads to the problem of distributing the aggregate cumulative revenue constraint across pricing groups. 
Taking into account the time value of money does not make this task easier; however, we can apply the heuristic described in Section~\ref{sec:distribution} --- distributing the difference between the constraint and expected revenue at current values of constants \(q^1, \ldots, q^k\) in proportion to the maximum possible increase in revenue relative to the expected revenue for each pricing group.
Alternatively, if the number of revenue constraints is small, various modifications of the Monte Carlo method can be applied.

\section{Numerical illustrations} \label{sec:sims} 
This section provides a visual demonstration of the basic algorithm. We first introduce an alternative approach for allocating revenue across pricing groups. Unlike the method in section~\ref{sec:distribution}, this new approach distributes aggregate cumulative revenue proportionally to the current revenue growth, rather than the difference between maximum possible revenue and expected revenue for each pricing group. Then we will demonstrate how the algorithm responds to changes in demand, which is critically important for pricing in real-world problems.

\subsection{Comparison of methods for solving the distribution problem}
In Section~\ref{sec:subsec:distrib_problem_statement}, we examined the problem of distribution of aggregate cumulative revenue across pricing groups, and in~\ref{sec:distribution}, we presented a possible solution. However, it is evident that this should not be considered the only viable approach. In this section, we discuss an alternative, also reasonable approach to this problem. We show that the previously presented strategy has an advantage over the method considered here.


It is assumed that during the sales period \(T=10\), RE objects from two pricing groups are sold: one-bedroom and two-bedroom apartments. At the times \([2, 4, 6, 8]\), constraints are set on the final sales of one-bedroom units at \(550\) and two-bedroom units at \(600\), as well as on aggregate cumulative revenue with values \([0, 80000, 90000, 0, 100000]\). We set the initial price for the first pricing group to be \(90\) and for the second --- \(100\). The demand for each pricing group is assumed to be linear with the following law:

for the first pricing group
\begin{equation*}
    \begin{cases}
        300, & \text{if} \; a^1 - b^1 \cdot p > 300,\\
        0, & \text{if} \; a^1 - b^1 \cdot p < 0, \\
        300 \cdot (a^1 - b^1 \cdot p), & \text{otherwise},
    \end{cases}
\end{equation*}

and for the second pricing group
\begin{equation*}
    \begin{cases}
        500, & \text{if} \; a^2 - b^2 \cdot p > 500,\\
        0, & \text{if} \; a^2 - b^2 \cdot p < 0, \\
        500 \cdot (a^2 - b^2 \cdot p), & \text{otherwise}.
    \end{cases}
\end{equation*}

The coefficients \(a^1, b^1\) are selected in such a way that the price range is \([20, 120]\), and the coefficients \(a^2, b^2\) are selected in such a way that the price variations lie within \([90, 110]\).

Recall that \(R^{*i}(\tau_{(1)}, \tau_{(2)})\) denotes the revenue for the \(i\)-th pricing group obtained through even absorption over the interval \([\tau_{(1)}, \tau_{(2)}]\), and \(R^i(\tau)\) represents the cumulative revenue up to time \(\tau\) from the sale of RE objects of the \(i\)-th pricing group.
Similarly, \(R^{*}(\tau_{(1)}, \tau_{(2)})\) denotes the aggregate cumulative revenue \(\sum_{i=1}^k R^{*i}(\tau_{(1)}, \tau_{(2)})\), and \(R(\tau)\) represents the aggregate cumulative revenue \(\sum_{i=1}^k R^i(\tau)\).

Now we describe another approach to solving the distribution problem. 
Let the difference between the aggregate cumulative revenue accumulated with the current price and that at even absorption be written as
\begin{equation*}
\Delta R(\tau_{(2)}) = R(\tau_{(2)}) - R^*(\tau_{(2)}, \tau_{(1)}).
\end{equation*}
The alternative approach then proposes the following revenue distribution across pricing groups:
\begin{equation*}
\Delta R^i(\tau_{(2)}) = \frac{R^i(\tau_{(2)})}{R^*(\tau_{(2)}, \tau_{(1)})} \Delta R(\tau_{(2)}), \quad i = 1, \ldots, k.
\end{equation*}

Then, the price \(p^i\) for RE objects in the \(i\)-th pricing group at time \(\tau_{(2)}\) is found from the equation:
\begin{equation*}
    p^i \cdot v^i(p^i) = \frac{1}{\tau_{(2)} - \tau_{(1)}} \Delta R^i(\tau_{(2)}).
\end{equation*}

\begin{figure}[!h]
    \centering
    \includegraphics[width=1\linewidth]{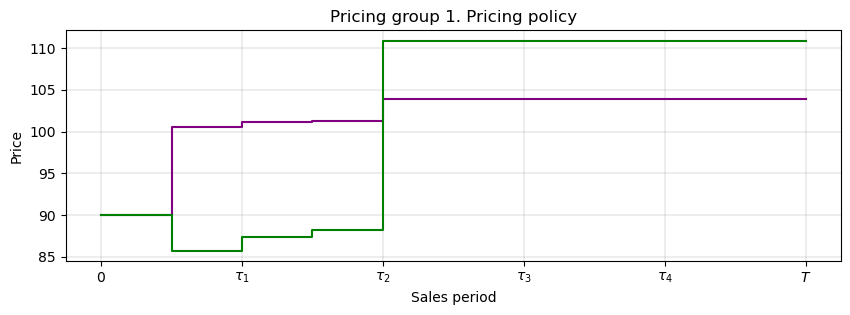}
    \caption{Comparison of pricing policies for RE objects in the first pricing group}
    \label{fig:adlas_price_1}
\end{figure}

\begin{figure}[!h]
    \centering
    \includegraphics[width=1\linewidth]{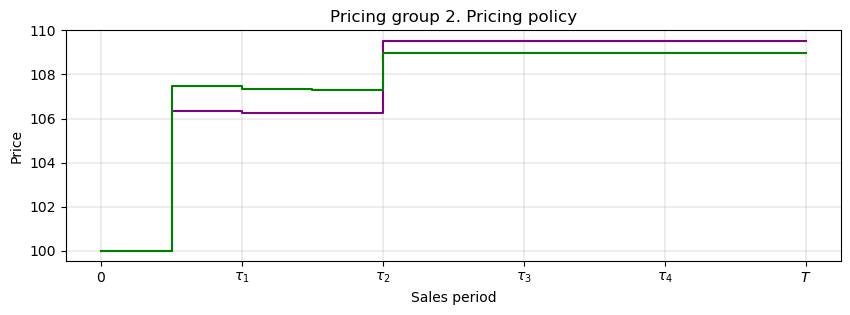}
    \caption{Comparison of pricing policies for RE objects in the second pricing group}
    \label{fig:adlas_price_2}
\end{figure}

Figures~\ref{fig:adlas_price_1} and \ref{fig:adlas_price_2} illustrate the pricing policies for the first and second pricing groups, respectively. Each graph has two curves: the purple line represents the pricing policy obtained using the method from Section~\ref{sec:distribution}, while the green line shows the pricing policy obtained using the alternative method.

\begin{figure}[!h]
    \centering
    \includegraphics[width=1\linewidth]{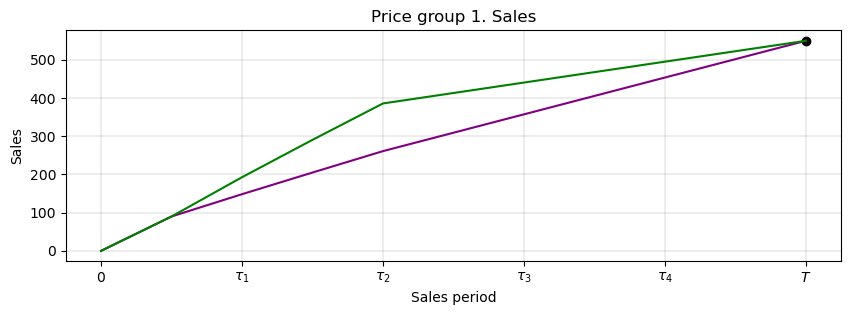}
    \caption{Comparison of sales curves for the first pricing group}
    \label{fig:adlas_sales_1}
\end{figure}

\begin{figure}[!h]
    \centering
    \includegraphics[width=1\linewidth]{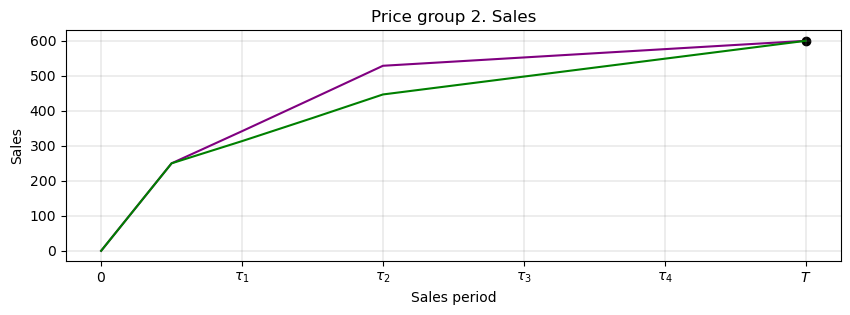}
    \caption{Comparison of sales curves for the second pricing group}
    \label{fig:adlas_sales_2}
\end{figure}

\begin{figure}[!h]
    \centering
    \includegraphics[width=1\linewidth]{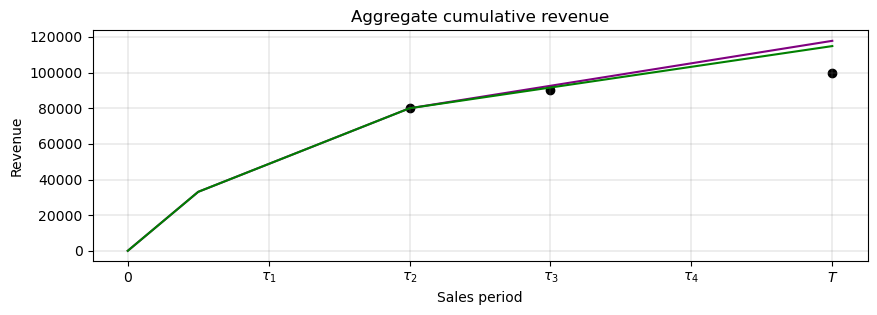}
    \caption{Aggregate cumulative revenue}
    \label{fig:adlas_revenue}
\end{figure}

Figures~\ref{fig:adlas_sales_1} and \ref{fig:adlas_sales_2} display curves showing cumulative sales for each pricing group. 
As shown in Figure~\ref{fig:adlas_revenue}, the curves are identical until time \(\tau_2\), but since the revenue constraint becomes burdensome at \(\tau_3\), the curves diverge after redistributing the difference between the goal and expected revenues at \(\tau_3\) across the pricing groups. Specifically, the cumulative revenue obtained with the method from Section~\ref{sec:distribution} is \(117839.78\), while the alternative method yields \(114889.51\). Thus, the distribution algorithm proposed in Section~\ref{sec:distribution} provides \(2.5\%\) higher revenue than the alternative algorithm.

\subsection{Visualization of algorithm's behaviour under changed demand}

Now let’s consider a situation where the demand function may change over the sales period \(T=10\). 
It is also assumed that at times \([2, 4, 6, 8]\) aggregate cumulative revenue constraints are set to be \([2500, 4000, 8000, 9500]\) and final sales constraints for one-bedroom apartments and for two-bedroom apartments are \(55\) and \(60\) respectively. 
We set the initial price for the first pricing group to be \(90\) and for the second --- \(100\).
For clarity, we will assume that (a) demand changes only once over the entire period, and (b) demand changes only for RE objects in one of the two pricing groups (in our example, for one-bedroom apartments). More precisely, let the demand functions for each pricing group \(i = 1, 2\) be given by:
\begin{equation*}
    v^1_t(p) =
    \begin{cases}
        20 \cdot (a^1 - b^1 \cdot p), & \text{if} \; t \in [0, \tau_3], \\
        4 \cdot (a^1 - b^1 \cdot p), & \text{if} \; t \in (\tau_3, T],
    \end{cases}
\end{equation*}
\begin{equation*}
    v^2(p) = 12 \cdot (a^2 - b^2 \cdot p).
\end{equation*}

\begin{figure}[h!]
    \centering
    \includegraphics[width=1\linewidth]{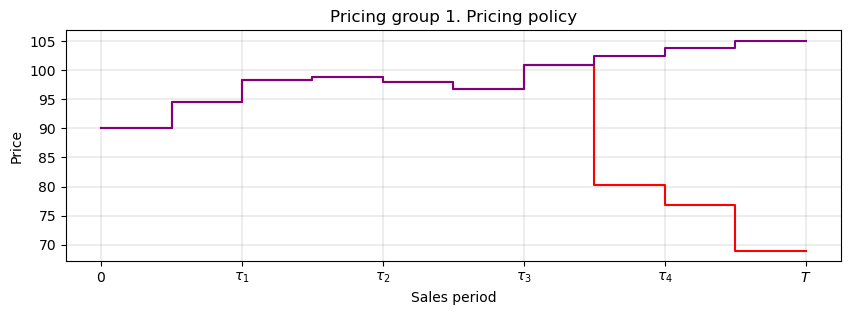}
    \caption{Pricing policy for pricing group 1}
    \label{fig:compare_price_1}
\end{figure}

\begin{figure}[h!]
    \centering
    \includegraphics[width=1\linewidth]{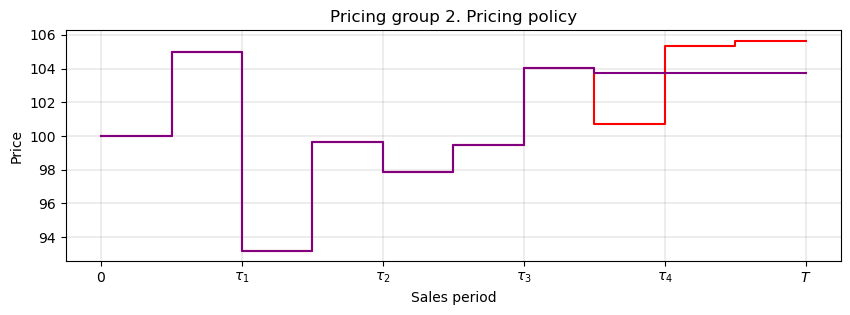}
    \caption{Pricing for pricing group 2}
    \label{fig:compare_price_2}
\end{figure}

For instance, a sharp decline in demand for real estate in the first pricing group was selected to illustrate that, even in such scenarios, the algorithm can consistently react to market changes and adjust the pricing strategy accordingly.
At the time when demand for the first pricing group changes, prices for RE objects in both pricing groups will be adjusted. Figures~\ref{fig:compare_price_1} and \ref{fig:compare_price_2} show the pricing policies for each pricing group. On each figure, at time \(\tau_3\), demand changes, and the purple line indicates the pricing policy if demand remained unchanged, while the red line represents the updated pricing policy. Note that although demand only changed for the first pricing group, pricing policies were adjusted for both. This is because each time the algorithm identifies a price change, it redistributes revenue across all pricing groups. There is also a practical reason for this adjustment: as demand for the first pricing group decreased, to meet the sales and revenue goals, prices needed to be adjusted both for one-bedroom apartments (lowering to improve conversion given the reduced demand) and for two-bedroom apartments to compensate for the decrease in demand.

\begin{figure}[!h]
    \centering
    \includegraphics[width=1\linewidth]{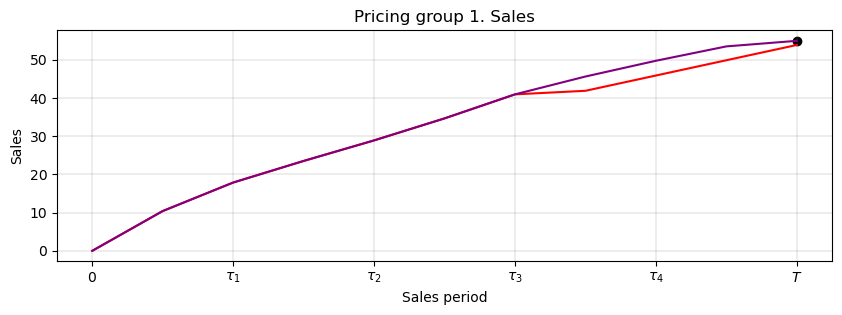}
    \caption{Cumulative sales for pricing group 1}
    \label{fig:compare_sales_1}
\end{figure}

\begin{figure}[h!]
    \centering
    \includegraphics[width=1\linewidth]{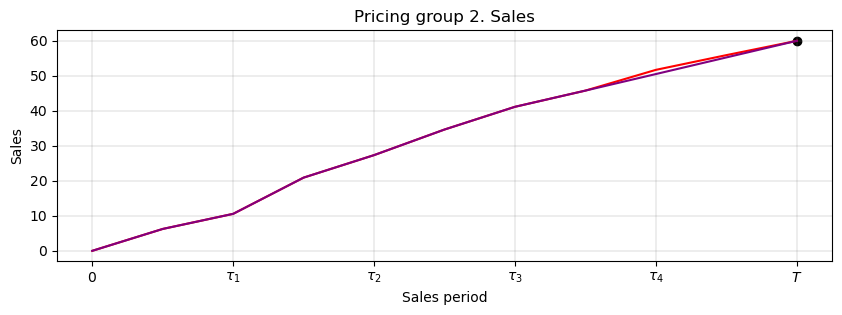}
    \caption{Cumulative sales for pricing group 2}
    \label{fig:compare_sales_2}
\end{figure}

Figures~\ref{fig:compare_sales_1} and \ref{fig:compare_sales_2} show the corresponding sales curves that illustrate how cumulative sales adjust to the decrease in demand. In Figure~\ref{fig:compare_sales_1}, it is noticeable that after the moment \(\tau_3\), when demand for the first pricing group declined, the sales curve also declined since demand was lower. However, by lowering prices for one-bedroom apartments, the algorithm managed to support sales volume with minimal losses and meet all constraints.

\begin{figure}[h!]
    \centering
    \includegraphics[width=1\linewidth]{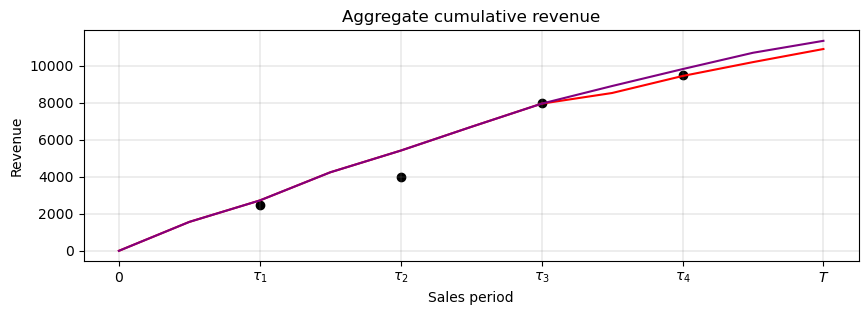}
    \caption{Aggregate cumulative revenue}
    \label{fig:compare_revenue}
\end{figure}

Figure~\ref{fig:compare_revenue} shows the aggregate cumulative revenue curve. 
Although by the end of the sales period, the revenue in the scenario with changing demand is lower than with constant demand, all financial constraints are satisfied, and the sales rate is maintained with minimal losses.
 
Finally, we present a visual illustration of how the algorithm redistributes revenue across pricing groups in response to changes in demand. 

\begin{figure}[!h]
    \centering
    \includegraphics[width=1\linewidth]{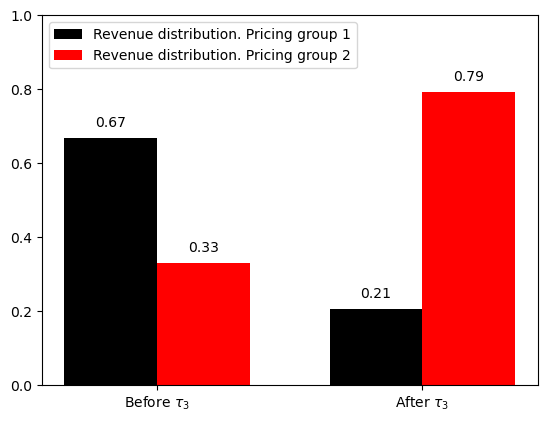}
    \caption{Revenue distribution between pricing groups}
    \label{fig:distrib_viz_bars}
\end{figure}

Figure~\ref{fig:distrib_viz_bars} shows a bar chart indicating how the portion of the revenue shortfall (in the algorithm, the difference between the maximum and expected revenue at time \(\tau_3\)) shifts with the change in demand. It can be seen that before \(\tau_3\), the revenue portion for the first pricing group was almost twice as large as for the second group, due to significantly higher demand for the first group. However, after demand for the first pricing group decreased, the importance of the second pricing group increased as its demand remained constant.

\section*{Conclusions}
This article examined dynamic pricing models that take into account multiple pricing groups of real estate. The basic pricing model was formulated and studied in the presence of several pricing groups.
The model was then modified to take into account the time value of money and the RE objects value increase as construction progresses.
The theoretical results regarding the general form of the pricing policy for both the basic model and its modification were obtained and proven.

The article also provides an algorithm for establishing a pricing policy for the basic model, as well as an algorithm for determining pricing policy under conditions of variable time value of money and the RE objects value as construction progresses.
A procedure for distributing aggregate cumulative revenue across the different pricing groups of real estate objects was presented.

Further improvements to the presented dynamic pricing model could include:
\begin{itemize}
\item Currently, the model requires cumulative sales and aggregate cumulative revenue to be greater than or equal to the sales and revenue goals.
However, in practice, it is common for market conditions to prevent the full achievement of goals, or for deviations from constraints to remain within reasonable limits.
Therefore, a significant improvement to the model could be the incorporation of "soft constraints", which, unlike the constraints discussed in the study, may be violated.

\item Incorporating reservations into the model could increase pricing accuracy.
Reservations provide preliminary demand and influence pricing decisions: if a certain amount of RE objects is reserved, the demand and price for the remaining objects can be adjusted accordingly.

\item Including installment plans and alternative payment options would broaden the appeal to buyers who may have difficulty paying the full price of the property upfront.
The model could account for changes in cash flow under an extended payment schedule and adjust prices for the remaining RE objects based on expected payments.
This approach creates additional competitive advantages in the market, particularly during periods of reduced purchasing power.
\end{itemize}

\bibliography{multiple_typologies}
\Addresses

\end{document}